# Some Enigmatic Aspects of the Early Universe III

C Sivaram and Kenath Arun

Indian Institute of Astrophysics, Bangalore- 560034

**Abstract:** In the previous parts of the discussion on the same topic, various aspects of the very early universe were discussed. We discussed how inclusion of large dark energy term compensates for the net gravity. Here the discussion is taken further including the effects of charge, magnetic fields and rotation. The role of large extra dimensions under the extreme initial conditions is discussed and possible connection with the cyclic brane theory is explored. We constrain various cosmic quantities like the net charge, number density of magnetic monopoles, primordial magnetic fields, size of the extra dimensions, etc.

The discussion in the previous parts was concerned with various aspects of the early universe. The scenario at the early stages of the collapsed universe is discussed including the effects of charge, magnetic fields and rotational effects. To do so we begin by noting that the exact solution for the metric of a particle of mass m in general relativity is given by the Schwarzschild solution as:

$$dS^2 = \left(1 - \frac{2GM}{rc^2}\right)c^2 dt^2 - \frac{dr^2}{\left(1 - \frac{2GM}{rc^2}\right)} - r^2 d\Omega^2 \qquad \ldots (1)$$

The solution for charged massive particle is given by the Reissner-Nordstrom solution as:[1]

$$dS^2 = \left(1 - \frac{2GM}{rc^2} + \frac{Ge^2}{c^4 r^2}\right)c^2 dt^2 - \frac{dr^2}{\left(1 - \frac{2GM}{rc^2} + \frac{Ge^2}{c^4 r^2}\right)} - r^2 d\Omega^2 \qquad \ldots (2)$$

From the above solution, for a massless charged particle, $g_{00} = 1 + \frac{Ge^2}{c^4 r^2} = 0$ would imply that $r = \sqrt{-G}\, e/c^2$, that is a naked singularity will arise. Due to this it is argued that a massless particle cannot have charge.[2]

But these zero rest mass particles will have a mass due to their electric charge.[3] The electric energy for a charge e of size r is given by: $E = \frac{e}{r^2}$

And the corresponding energy density is given by: $\frac{E^2}{8\pi} = \frac{e^2}{8\pi r^4}$

The total energy density associated with this charge is: $\varepsilon = \int_r^\infty \frac{e^2}{8\pi r^4} 4\pi r^2 dr = \frac{e^2}{2r}$

Therefore the mass due to the electric charge is given by: $M = \frac{e^2}{2rc^2}$

From this we have: (cf. equation (2))

$$g_{00} = 1 - \frac{2G}{rc^2}\left(\frac{e^2}{2rc^2}\right) + \frac{Ge^2}{c^4 r^2} \qquad \ldots (3)$$

The last two terms cancels and hence we have in this case: $g_{00} = 1$, that is it gives a flat space-time. Thus if a particle's mass is only due to its electric charge (like perhaps for the classical electron), then it will not bend the space-time around it!

Including the dark energy term, assumed to be just the cosmological constant, $\Lambda$,[4] the solution is given by the Reissner-Nordstrom de-sitter solution (also called the Kottler metric) as:

$$dS^2 = \left(1 - \frac{2GM}{rc^2} + \frac{Ge^2}{c^4 r^2} - \frac{\Lambda r^2}{3}\right) c^2 dt^2 - \frac{dr^2}{\left(1 - \frac{2GM}{rc^2} + \frac{Ge^2}{c^4 r^2} - \frac{\Lambda r^2}{3}\right)} - r^2 d\Omega^2 \quad \dots (4)$$

Where,

$$g_{00} = 1 - \frac{2GM}{rc^2} + \frac{Ge^2}{c^4 r^2} - \frac{\Lambda r^2}{3} \quad \dots (5)$$

If the cosmological constant term compensates for the charge in equation (5), then as for the electron $\frac{2GM}{rc^2} \ll 1$, we have, $r = \left(\frac{3Ge^2}{c^4 \Lambda}\right)^{1/4}$.

That is, the last two terms in equation (5) become equal at this value of $r$ (where $\Lambda = 10^{-56} cm^{-2}$). For an electron this works out to be of the order of $10^{-3} cm$.

This suggests that the electrostatic energy density around a single electron becomes comparable to the cosmological constant energy density at sub-millimetre distances. This could be testable in an experiment involving single electrons in devices like an ion trap.

If we consider the effects of quantum vacuum fluctuation term, where energy density is given by $\frac{\beta \hbar c}{r^4}$, where $\beta \sim 1$ (like in the case of Casimir force), the metric then becomes

$$g_{00} = 1 - \frac{2GM}{rc^2} + \frac{\beta G \hbar c}{c^4 r^2} - \frac{\Lambda r^2}{3}$$

As in the earlier case, the last two terms become comparable for a radial distance given by:

$$r = \left(\frac{3G\hbar}{c^3 \Lambda}\right)^{1/4} \quad \dots (6)$$

Which again works out to be $\sim 3\times 10^{-3} \, cm$.

For a spinning body we get a similar expression as equation (6), that is,

$$r = \left(\frac{3GJ}{c^3 \Lambda}\right)^{1/4} \qquad \ldots (7)$$

where $J$ is the angular momentum.

From equation (7) we see that:

$$\frac{J}{r^4} = \frac{c^3}{3G}\Lambda \qquad \ldots (8)$$

which is a constant. We have already shown (for a very large range of structures in the universe) that the following relation holds. That is:[5, 6]

$$\frac{M}{r^2} = \frac{c^2}{G}\sqrt{\Lambda} \qquad \ldots (9)$$

Which is again a constant. Therefore equations (8) and (9) together imply that $J \propto M^2$. This is consistent with what is actually observed for a wide range of astronomical structures.[7, 8]

For the cosmological case (where the total action is $\sim J = 10^{120}\hbar$), $r$ then corresponds to the size of the universe, that is:

$$r = \left(\frac{3GJ}{c^3 \Lambda}\right)^{1/4} = (3\times 10^{-3}\, cm)(10^{120})^{1/4} \sim 10^{28}\, cm \qquad \ldots (10)$$

In the case of a typical galaxy, $J = 10^{100}\hbar$, this implies:

$$r = \left(\frac{3GJ}{c^3 \Lambda}\right)^{1/4} = (3\times 10^{-3}\, cm)(10^{100})^{1/4} \sim 3\times 10^{22}\, cm \qquad \ldots (11)$$

In the case of a star, $J = 10^{76}\hbar \Rightarrow r \sim 10^{16}\, cm$, which is the expected size of the nebula (interstellar cloud) from which the star formed.[8, 9, 10]

We can see that the dark energy term goes as $1/r^4$ in the metric, so for the recollapsed universe as a whole at $r \sim 10^{-3}\, cm$, the cosmological constant term is of the order of $\Lambda = \Lambda_{Pl} \sim 10^{66}\, cm^{-2}$, (that is in the final stage of collapse) and this is consistent with earlier results.[11, 12]

At the present epoch, according to the above relation:

$$\Lambda = \Lambda_{Pl}\left(\frac{3\times 10^{-3}}{10^{28}}\right)^4 \sim 10^{-56}\, cm^{-2} \qquad \ldots (12)$$

Which corresponds to the observed value![13]

But $r$ scales with time as: $r \propto t^{1/2}$ in the radiation era, this implies that $\Lambda \propto \dfrac{1}{t^2}$

At the present epoch, that is $10^{18}\,s$ (Hubble time), the dark energy term is given by:

$$\Lambda = \Lambda_{Pl}\left(\frac{10^{-43}}{10^{18}}\right)^2 \sim 10^{-56}\, cm^{-2} \qquad \ldots (13)$$

This is again in agreement with what is observed. We can think of what the implications for a closed universe, recollapsing to a finite size of $r \approx 10^{-3}\, cm$ are, as implied by the above equations and also elaborated in earlier parts.[11, 12]

We can try and interpret in what follows. We note that in many current models of unification of interactions, there are large extra dimensions of sub-millimetre ($r \approx 10^{-3}\, cm$) size.[14]

The zero point energy (ZPE) fluctuations at the boundary between our 3-space and the extra dimension can give rise to a vacuum energy density in our universe.

$L^{-1}$ gives a wave number (k) cut-off so that vacuum energy density propagating into our universe gives rise to an effective cosmological constant $\Lambda$ obtained as:[15]

$$\Lambda \approx \frac{L_{Pl}^2}{L_{EW}^4\left(L_{EW}/L_{Pl}\right)^{8/n}} \qquad \ldots (14)$$

Where $n$ is the number of extra dimensions and $L_{Pl}$ is the familiar Planck length. For $n = 2$, we have:[15, 16]

$$\Lambda \approx \frac{L_{Pl}^6}{L_{EW}^8} = \frac{\hbar^7 G^3}{G_F^4 c^5} \sim 10^{-56}\, cm^{-2} \qquad \ldots (15)$$

Where, $L_{Pl} = \left(\hbar G/c^3\right)^{1/2}$, $L_{EW} = \left(G_F/\hbar c\right)^{1/2} = 7\times 10^{-17}\, cm$ is the beta decay length and $G_F = 1.5\times 10^{-49}\, erg\, cm^3$ is the Fermi constant.

This precisely gives the size of the extra dimensions as $L \approx 0.01 cm$. This value of L can also be understood as the Casimir energy between two branes (separated by L) balancing the repulsive cosmic density:[16, 17]

$$L \approx \left( \frac{8\pi G \hbar c}{240 \Lambda c^4} \right)^{1/4} \sim 0.01 cm \qquad \ldots (16)$$

Size of the extra dimension L, and their number $n$, are independent parameters but in order to achieve equality of strength of gravity and electroweak force at $L_{EW} \sim 10^{-19} m$, they become constrained by $L = 10^{(32/n)-19}$. This is in the spirit of models with large extra dimensions.

For large $n$, strength of gravity grows very rapidly at microscopic length and becomes comparable to the electroweak force at $\sim 10^{-19} m$.[14, 18]

The size of the extra dimensions is given by:

$$R_C = \frac{\hbar c}{M_W c^2} \left( \frac{M_{Pl}}{M_{EW}} \right)^{2/n} \sim L_{EW}^n \times 10^{32} \qquad \ldots (17)$$

For $n = 1$, the size of the extra dimension is given by, $R_C \sim 10^{-17} \times 10^{32} = 10^{15} cm$, which is about the size of the solar system and there are no observed deviations from the usual 4-dimensional space in this scale, hence $n = 1$ is not a possible number for a large extra dimension.

For $n = 2$, the size of the extra dimension is given by, $R_C \sim 10^{-3} cm$, which is testable and is indeed being explored in numerous experiments.[19]

So if the universe recollapses to $L \sim 10^{-3} cm$, it can tunnel into this extra dimension given by equation (17) to another space-time and reexpand.

The force per unit area or the energy density between two branes given by:[18, 20]

$$\frac{F}{A} = \varepsilon = \frac{\hbar c}{240 \pi d^4} \qquad \ldots (18)$$

In terms of the cosmological constant it is given by: $\varepsilon = \dfrac{\Lambda c^4}{8\pi G}$, where we have effectively:

$$\Lambda = \dfrac{\hbar^7 G^3}{G_F^4 c^5} \qquad \ldots (19)$$

Therefore the energy density is given by:

$$\varepsilon = \dfrac{\hbar^7 G^2}{8\pi G_F^4 c} \qquad \ldots (20)$$

This has the value of $\sim 10^{-8}\, ergs/cm^3$, precisely what is observed from the dark energy.

According to general relativity, the maximum force is $F_{max} = \dfrac{GM^2}{R_{min}^2}$, where, $R_{min} \approx \dfrac{GM}{c^2}$.

This implies that $F_{max} = \dfrac{c^4}{G}$.

Therefore the maximum area to which it can expand, under the pressure (energy density) given by equation (20) is given by maximum force/energy density. That is given by

$$A_{max} = \dfrac{8\pi G_F^4 c \cdot \dfrac{c^4}{G}}{\hbar^7 G^2} = \dfrac{8\pi G_F^4 c^5}{\hbar^7 G^3} \approx 10^{56}\, cm^2 \qquad \ldots (21)$$

And the corresponding size is given by $R = 10^{28}\, cm$ which is the Hubble radius.
So we have a possible scenario to explain the present scale of the expanding universe.

From equation (20), the energy density is given by: $\varepsilon = \dfrac{\hbar^7 G^2}{8\pi G_F^4 c} \sim 10^{-8}\, ergs/cm^3$

The energy associated with the total volume is therefore given by:

$$E_\Lambda = \dfrac{\hbar^7 G^2}{8\pi G_F^4 c} \times 2\pi^2 \Lambda^{-3/2} = \dfrac{\pi G_F^2 c^{13/2}}{4\hbar^{7/2} G^{5/2}} \sim 2 \times 10^{77}\, ergs \qquad \ldots (22)$$

Where $2\pi^2 \Lambda^{-3/2}$ is the volume. This energy quantifies the total dark energy in the universe. The corresponding mass is given by:

$$M_\Lambda = \dfrac{E_\Lambda}{c^2} = \dfrac{\pi G_F^2 c^{9/2}}{4\hbar^{7/2} G^{5/2}} \sim 2 \times 10^{56}\, g \qquad \ldots (23)$$

The maximum power associated with the baryonic matter is given by general relativity as $c^5/G$. This can be seen as follows.

The total energy released is $\sim \frac{GM^2}{R}$, and the total power is given by:

$$P \approx \frac{GM^2}{Rt} \quad \ldots (24)$$

In general relativity, the smallest time scale (or smallest length scale) associated with a given mass is $t_{min} \approx \frac{GM}{c^3}$ (for $R$ corresponding to the gravitational radius of $R \approx \frac{GM}{c^2}$).

This gives the maximal power as:

$$P \approx \frac{GM^2}{Rt} \approx \frac{c^5}{G} \quad \ldots (25)$$

(Substituting the minimal values for $R$ and $t$ as given above)

So if we have a given set of objects of total baryonic mass $(\Sigma M)$, generating radiation energy we can write for the combined maximal Eddington luminosity as:[21, 22]

$$\frac{4\pi G(\Sigma M)cm_P}{\sigma_T} = \frac{c^5}{G} \quad \ldots (26)$$

Where, $m_P$ is the proton mass and $\sigma_T$ is the Thomson cross section and $\Sigma M = M_b$ gives the total baryonic mass, which generates the radiation luminosity.

This gives:[21]

$$M_b = \frac{\sigma_T c^4}{4\pi G^2 m_P} \sim 9 \times 10^{54} \, g \quad \ldots (27)$$

From equations (23) and (27) we can see that:

$$\frac{M_\Lambda}{M_b} \sim 25 \quad \ldots (28)$$

This is consistent with observations.[23]

The maximum area to which this mass can expand is given by equation (21) as $A_{max} = \frac{8\pi G_F^4 c^5}{\hbar^7 G^3}$. And the mass by area is given by:

$$\frac{M_\Lambda}{A} = \frac{\pi G_F^2 c^{9/2}}{4\hbar^{7/2}G^{5/2}} \times \frac{\hbar^7 G^3}{8\pi G_F^4 c^5} = \frac{\hbar^{7/2}G^{1/2}}{32 G_F^2 c^{1/2}} \sim 1 g/cm^2 \qquad \ldots (29)$$

This is consistent with the scaling relations for the mass as obtained for a wide range of self-similar structures in the universe, right from the scale of the electron to that of the entire universe.[5, 6, 23]

We can also arrive at the maximum magnetic field possible at the present epoch based on the results obtained above. The maximum energy density due to the magnetic field at the Planck epoch is given by:

$$\frac{B^2}{8\pi} = \rho_{Pl} = \frac{\left(\frac{\hbar c^5}{G}\right)^{1/2}}{\frac{4}{3}\pi\left(\frac{\hbar G}{c^3}\right)^{3/2}} \approx 10^{114} \, ergs/cm^3 \qquad \ldots (30)$$

The corresponding maximum field is then given by:[24, 25]

$$B_{max} = \left(8\pi \times 10^{114}\right)^{1/2} \approx 10^{57} G \qquad \ldots (31)$$

Since the flux ($BR^2$) is conserved we can calculate the magnetic field for the present epoch. That is:

$$B_0 R_0^2 = B_{max} R_{min}^2 \qquad \ldots (32)$$

Where $R_{min} = 10^{-3} cm$ is the size corresponding at the Planck epoch.

The magnetic field at the present epoch is then given by:

$$B_0 = 10^{57}\left(\frac{R_{min}}{R_0}\right)^2 \approx 10^{-6} G \qquad \ldots (33)$$

This is the maximum possible magnetic field at the present epoch. But the microwave background sets constraints on the energy density due to this magnetic field. At BBN era corresponding to time of 1 second, the size of the universe and the temperature are of the order of $10^{18} cm$ and $10^{10} K$.

The corresponding energy density due to the radiation is given by:

$$\rho_{rad} = aT^4 \approx 10^{26} \, ergs/cm^3 \qquad \ldots (34)$$

The microwave background sets a constraint that the energy density due to the magnetic field be less than 5% of that given by equation (34).[23] That is

$$\rho_{mag} = \frac{B^2}{8\pi} \approx 10^{24} \, ergs/cm^3 \qquad \ldots (35)$$

The maximum magnetic field of $10^{11} G$ corresponds to temperature of $10^{10} K$ and size of $10^{18} cm$. Since the flux is conserved, the field at the present epoch is given by:

$$B_0 = 10^{11} \left(\frac{10^{18}}{10^{28}}\right)^2 \approx 10^{-9} G \qquad \ldots (36)$$

This is consistent with the observed Faraday rotation, in extragalactic structures.[25]

We have seen from equation (31) that the maximum magnetic field at the Planck epoch is of the order of $\sim 10^{57} G$ and the corresponding flux is given by:

$$BR_{min}^2 \approx 10^{52} \, Gcm^2 \qquad \ldots (37)$$

Where $R_{min} = 10^{-3} cm$ is the size corresponding at the Planck epoch.
The unit quantum of flux is given by:[26]

$$\frac{\hbar c}{2e} \sim 10^{-8} \qquad \ldots (38)$$

From equations (37) and (38) we see that there are $10^{60}$ units of flux quanta. This can be interpreted as the maximum number of monopoles. This can be explained as follows. From equation (33) we see that the maximum magnetic field allowed at the present epoch is $\sim 10^{-6} G$, but from the consideration of conservation of flux we arrived at the field at the present epoch as $\sim 10^{-9} G$ (equation (36)) which is consistent with observation. This could be because the remaining energy is trapped in these monopoles.

Therefore the $10^{60}$ units of flux quanta can be interpreted as the maximum number of monopoles. Since the flux (number of monopoles) is conserved, there are $10^{60}$ monopoles in the present volume of $2\pi^2 \Lambda^{-3/2} \sim 2 \times 10^{85} cm^3$.
That is the monopole density at the present epoch is:

$$\sim \frac{10^{60}}{2 \times 10^{85}} \sim 5 \times 10^{-26} /cc \qquad \ldots (39)$$

Or the monopole flux is given by:

$$\sim 5 \times 10^{-26} \times \frac{c}{4} \approx 10^{-16} \, cm^{-2} s^{-1} \qquad \ldots (40)$$

This matches with the bound on the monopole flux of $10^{-16} \, cm^{-2} s^{-1}$ set by Parker from independent considerations.[27]

We had for the single electron charge, the radius as:

$$r = \left(\frac{3Ge^2}{c^4 \Lambda}\right)^{1/4} \approx 10^{-3} \, cm \qquad \ldots (41)$$

For electron charge $e$ and $\Lambda \approx 10^{-56} \, cm^{-2}$ (that is the present value).

In the initial stage, we had for the whole universe, $\Lambda = \Lambda_{Pl}$ and the corresponding size works out to be:

$$r = \left(\frac{3Gq^2}{c^4 \Lambda_{Pl}}\right)^{1/4} \approx 10^{-3} \, cm \qquad \ldots (42)$$

Comparing equations (41) and (42), it follows that since $r$ is of comparable magnitude, $q^2 \approx 10^{120} e^2$, giving:

$$q \approx 10^{60} e \qquad \ldots (43)$$

This gives the maximal value of the total electric charge we can have. As the radius encompasses $10^{90}$ particles, this would imply that net charge (in electron unit) is 1 part in $10^{30}$.[28, 29]

So it appears interesting that the upper limit of the number of magnetic flux quanta $(\sim 10^{60})$ is also the same as the upper limit on the number of electric charge. This suggests a electric-magnetic charge duality in the spirit of Dirac (who explained quantification of electric charge on the basis of existence of magnetic monopole via the quantisation condition $eg \approx \hbar c$, where $g$ is the magnetic charge).[30, 31, 32]